\documentclass[twocolumn,english,aps,prl,superscriptaddress]{revtex4-1}
\usepackage[T1]{fontenc}
\usepackage[latin9]{inputenc}
\usepackage{amsmath}
\usepackage{amssymb}
\usepackage{graphicx}
\usepackage{esint}

\makeatletter
 
 \@ifundefined{textcolor}{}
 {%
   \definecolor{BLACK}{gray}{0}
   \definecolor{WHITE}{gray}{1}
   \definecolor{RED}{rgb}{1,0,0}
   \definecolor{GREEN}{rgb}{0,1,0}
   \definecolor{BLUE}{rgb}{0,0,1}
   \definecolor{CYAN}{cmyk}{1,0,0,0}
   \definecolor{MAGENTA}{cmyk}{0,1,0,0}
   \definecolor{YELLOW}{cmyk}{0,0,1,0}
 }

\@ifundefined{definecolor}{\usepackage{color}}{}

\makeatother

\usepackage{babel}
\begin{document}

\preprint{This line only printed with preprint option}

\title{Green's functions from real-time bold-line Monte Carlo: spectral
properties of the nonequilibrium Anderson impurity model}

\author{Guy Cohen}

\affiliation{Department of Chemistry, Columbia University, New York, New York
10027, U.S.A.}

\affiliation{Department of Physics, Columbia University, New York, New York 10027,
U.S.A.}

\author{Emanuel Gull}

\affiliation{Department of Physics, University of Michigan, Ann Arbor, MI 48109,
U.S.A.}

\author{David R. Reichman}

\affiliation{Department of Chemistry, Columbia University, New York, New York
10027, U.S.A.}

\author{Andrew J. Millis}

\affiliation{Department of Physics, Columbia University, New York, New York 10027,
U.S.A.}
\begin{abstract}
The nonequilibrium spectral properties of the Anderson impurity model
with a chemical potential bias are investigated within a numerically
exact real time quantum Monte Carlo formalism. The two-time correlation
function is computed in a form suitable for nonequilibrium dynamical
mean field calculations. Additionally, the evolution of the model's
spectral properties are simulated in an alternative representation,
defined by a hypothetical but experimentally realizable weakly coupled
auxiliary lead. The voltage splitting of the Kondo peak is confirmed
and the dynamics of its formation after a coupling or gate quench
are studied. This representation is shown to contain additional information
about the dot's population dynamics. Further, we show that the voltage-dependent
differential conductance gives a reasonable qualitative estimate of
the equilibrium spectral function, but significant qualitative differences
are found including incorrect trends and spurious temperature dependent
effects. 
\end{abstract}
\maketitle
The nonequilibrium physics of strongly correlated systems is a fundamental
issue at the cutting edge of research in condensed matter physics.
Out-of-equilibrium processes can be manipulated and studied in cold
atomic gases \cite{mandel_coherent_2003,kinoshita_quantum_2006,hofferberth_non-equilibrium_2007}
or by using ultrafast spectroscopy \cite{iwai_ultrafast_2003,perfetti_time_2006},
and are relevant for the understanding of phenomena ranging from the
behavior of atoms \cite{madhavan_tunneling_1998}, molecules \cite{zhao_controlling_2005}
and nanocrystals \cite{mocatta_heavily_2011} adsorbed on surfaces
to transport in molecular electronic devices \cite{park_coulomb_2002,heath_molecular_2003}.
The problem is theoretically challenging because the strong correlations
render perturbative techniques inapplicable while the nonequilibrium
aspects preclude the use of most standard statistical-mechanics techniques.
The principal methods rely on real time propagation from some initial
condition and are limited in the times which can be accessed. For
steady state a numerically exact description in terms of Matsubara
voltages can bypass time propagation, but becomes biased by the need
to perform analytical continuation \cite{han_imaginary-time_2007,dirks_continuous-time_2010}.
Direct equation of motion techniques are sometimes applicable where
initial correlations can be neglected \cite{jin_exact_2008}, but
at computational costs similar to direct propagation. The theoretical
challenges become particularly acute when one is interested in steady
state correlation functions: converged results require propagation
to times long enough so that steady state is reached, and beyond that
to the times needed to define the correlation function.

One simplifying aspect of many interesting cases is that the important
many-body correlations may be taken to be localized in space, either
by the physical situation (for example a quantum dot where the interactions
are confined to the region of the dot and the leads may be taken to
be noninteracting) or by a theoretical approximation such as dynamical
mean field theory (DMFT) which may be formulated both in \cite{georges_dynamical_1996,georges_strongly_2004}
and out of \cite{freericks_nonequilibrium_2006,freericks_steady-state_2006}
equilibrium and provides an approximate solution of the properties
of a spatially infinite system in terms of the solution of a quantum
impurity model.

A crucial bottleneck in the applications of DMFT to the nonequilibrium
situation has been the lack of impurity solvers which can access the
long time behavior. In particular, nonequilibrium DMFT requires the
evaluation of the dynamical electron propagator, a two-time correlation
function. Recent years have seen the development of several controlled
nonequilibrium impurity solvers with this capability, including work
based on interaction expansion Monte Carlo \cite{eckstein_thermalization_2009},
exact diagonalization \cite{arrigoni_nonequilibrium_2013,gramsch_2013}
and hierarchical equation of motion techniques \cite{li_hierarchical_2012,wang_hierarchical_2013,hou_quantum_2013}.
These approaches have provided important insights into the physics
of strongly correlated systems out of equilibrium, but all carry intrinsic
limitations and are viable only in particular parameter regimes. Monte
Carlo methods are restricted by the dynamical sign problem to short
propagation times, making it difficult to obtain high-resolution spectral
information or access the nonequilibrium steady state \cite{dirks_extracting_2013}.
On the other hand, the exact diagonalization and equation of motion
methods have a very unfavorable computational scaling because the
the spectral structure of the noninteracting baths must be represented
by a small number of degrees of freedom. A general and unbiased computational
scheme capable of representing spectral data at the level required
for comparison to experiment or for general DMFT applications remains
sorely needed.

Recently, a method for extracting numerically exact spectral information
and correlation functions from real time bold-line \cite{prokofev_bold_2007,prokofev_bold_2008}
continuous time Monte Carlo (bold-CTQMC) \cite{gull_bold-line_2010,gull_numerically_2011}
has been put forth \cite{cohen_unpublished} which largely circumvents
many of the limitations of previous real-time Monte Carlo methods.
The method can access substantially longer times than were previously
accessible, and in combination with memory function methods \cite{cohen_memory_2011-1,cohen_generalized_2013,wilner_multiple_2013}
has been shown to enable the computation of single-time observables
such as the magnetization density out to unprecedentedly long times
\cite{cohen_numerically_2013}. In this Letter we show that the new
bold-line methods enable the calculation of the steady-state nonequilibrium
two-time electron Green's function and lead to new insights into the
evolution of the system towards steady state for the prototypical
example of the nonequilibrium Anderson impurity model. We follow the
formation of the Kondo peak after a gate quench, showing how the electron
spectral function evolves to its steady state value; demonstrate the
long-suspected voltage splitting of the Kondo resonance in the presence
of a bias voltage \cite{meir_low-temperature_1993,de_franceschi_out--equilibrium_2002,leturcq_probing_2005,leturcq_kondo_2006,shah_2006,fritsch_nonequilibrium_2010,pletyukhov_nonequilibrium_2012};
and establish that the current-voltage characteristic of a quantum
dot provides an inaccurate representation of the many-body density
of states. The impurity solver described here works in a manner practical
for the needs of nonequilibrium DMFT. In particular, the computational
complexity of our approach is independent of both the dot--bath coupling
density and the final spectral resolution desired.

\begin{figure}
\includegraphics[width=8.6cm]{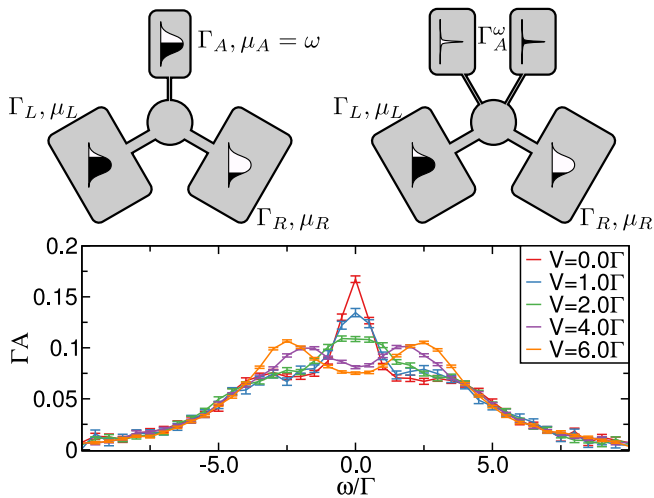}\caption{The experimental auxiliary lead setup (top left) and the double probe
scheme (top right) are illustrated. Below, the steady state spectral
function $A\left(\omega\right)$ is shown at several voltages. The
results are obtained from bold-CTQMC using the double probe auxiliary
lead formalism at $\Gamma t=10$. Error bars estimate statistical
Monte Carlo errors.\label{fig:voltage-effect}}
\end{figure}

We have used the bold-line methods \cite{gull_bold-line_2010,gull_numerically_2011}
to directly evaluate two time correlation functions but we find that
more accurate and efficient access to the steady state spectral function
may be obtained from a variant of an insightful idea originally proposed
as an experimental configuration for probing transport in quantum
dots \cite{lebanon_measuring_2001,sun_kondo_2001}. In its original
form the idea was to relate the spectral function to the voltage dependence
of a current flowing through a single additional weakly coupled auxiliary
lead $A$ (Fig.~\ref{fig:voltage-effect}, top left):

\begin{equation}
A_{\mathrm{aux}}(V_{A},t)=\lim_{\Gamma_{A}\rightarrow0}-\frac{1}{\Gamma_{A}\pi}\frac{dI_{A}(t)}{dV_{A}}.
\end{equation}
As $t$ approaches infinity while the auxiliary lead is kept at a
fixed chemical potential $V_{A}$, $A_{\mathrm{aux}}\left(V_{A},t\right)$
becomes time-independent and approaches $A\left(\omega=V_{A}\right)\equiv-\frac{1}{\pi}\Im\left\{ G^{r}\left(\omega=V_{A}\right)\right\} $.
We find \cite{cohen_unpublished} that a theoretically more convenient
(although experimentally impractical) representation may be achieved
by considering the current $I$ flowing between two auxiliary leads
(Fig~\ref{fig:voltage-effect}, top right) which are weakly coupled
to the systems only at a predefined frequency $\omega^{\prime}$ {[}$\Gamma_{A}=\eta\delta\left(\omega-\omega^{\prime}\right)${]}
with $\eta$ much less than the typical physical coupling $\Gamma$
to the principal leads. We take one of the leads to be full (f: chemical
potential much higher than any relevant scale) and one to be empty
(e: chemical potential much lower than any relevant scale). Then 
\begin{equation}
A_{\mathrm{aux}}\left(\omega,t\right)=\lim_{\eta\rightarrow0}-\frac{2h}{e\pi\eta}\left[I_{A}^{\mathrm{f}}\left(\omega,t\right)-I_{A}^{\mathrm{e}}\left(\omega,t\right)\right].
\end{equation}
In addition to its computational advantages, this formalism provides
physical insight into the evolution of dot properties after a quench.
At any given time, the full lead injects electrons into the system
at frequency $\omega$ and at a rate of $-I_{A}^{\mathrm{f}}\left(\omega,t\right)$,
and should thus (neglecting the response properties of the auxiliary
lead itself) be proportional to the density of electronic excitations
at this frequency and time; similarly, $I_{A}^{e}\left(\omega,t\right)$
probes the density of hole excitations. Experimentally, one would
only have access to $A_{\mathrm{aux}}$, which is proportional to
the total (electron+hole) excitation density. In equilibrium or in
steady state outside the bias window (up to $\sim k_{B}T$) clearly
only the empty or full probe contributes and excitations can be distinguished
by type. For comparison, if $A(t)$ is obtained only for a finite
time interval, its Fourier transform yields only a discrete set of
energies approximating $A(\omega)$. Since $A_{\mathrm{aux}}\left(\omega,t\right)$
provides frequency-rich information at all times, and since (unlike
the two-time correlation function) an experimental pathway for directly
measuring it has been suggested, we suggest that it is an interesting
and potentially useful quantity to explore in its own right.

The model we treat consists of an Anderson impurity \cite{anderson_localized_1961}
coupled to two leads held at different chemical potentials (upper
panels, Fig.~\ref{fig:voltage-effect}). Physical realizations include
transport in molecular junctions and scanning microscopy studies of
adsorbed atoms. However, we emphasize that the method is equally applicable
to other nonequilibrium situations including Hamiltonians with explicit
time dependence arising in irradiated quantum dots and in the dynamical
mean field analysis of pump-probe experiments. Setting $\hbar=e=1$,
the Anderson model Hamiltonian is 
\begin{align}
H & =H_{D}+H_{B}+V,\label{eq:hamiltonian}\\
H_{D} & =\sum_{\sigma\in\left\{ \uparrow,\downarrow\right\} }\varepsilon_{\sigma}d_{\sigma}^{\dagger}d_{\sigma}+Un_{\uparrow}n_{\downarrow},\\
H_{B} & =\sum_{a=L,R\sigma k}\varepsilon_{\sigma k}a_{a\sigma k}^{\dagger}a_{a\sigma k},\\
V & =\sum_{a\sigma k}\left(V_{a\sigma k}a_{a\sigma k}^{\dagger}d_{\sigma}+\mathrm{H.C.}\right).
\end{align}
Here, $d_{\sigma}^{\dagger}\left(d_{\sigma}\right)$ operators create
(destroy) electrons with spin $\sigma=\pm\frac{1}{2}$ and energy
$\varepsilon_{\sigma}$ on the dot; $a_{a\sigma k}^{\dagger}\left(a_{\sigma k}\right)$
operators create (destroy) electrons with spin $\sigma$ and energy
$\varepsilon_{\sigma k}$ in the left ($a=L$) or right ($a=R$) lead,
where the indices $k$ enumerate levels; and the $V_{a\sigma k}$
define the dot-lead hybridization (an analogous definition applies
to the auxiliary lead). The lead dispersions and the coupling strengths
are determined by a coupling density $\Gamma_{L/R}\left(\omega\right)=2\pi\sum_{k\in L/R}V_{\sigma k}^{*}V_{\sigma k}\delta\left(\omega-\varepsilon_{k}\right)$.
In the rest of this paper we will take the $\Gamma$ to be identical
for the two leads and spins (this is done for convenience and is by
no means a limitation of the method). We choose a flat, soft edged
coupling density $\Gamma_{L/R}\left(\omega\right)=\frac{\Gamma/2}{\left(1+e^{\nu\left(\omega-\Omega_{c}\right)}\right)\left(1+e^{-\nu\left(\omega+\Omega_{c}\right)}\right)}$,
and in order to keep the discussion simple all results shown are at
an interaction of $U=6\Gamma$, an inverse temperature of $\beta\Gamma=3$,
a bandwidth of $\Omega_{c}=10\Gamma$ and an inverse band edge width
of $\nu=10\Gamma^{-1}$ (except where stated otherwise). The system
is expected to have a Kondo temperature of $\sim0.2\Gamma$. We also
hold the chemical potentials in the two leads at a symmetrically applied
bias $\mu_{L/R}=\pm\frac{V}{2}$ (we note in passing that the main
limitation of the method is in accessing low temperatures\cite{gull_bold-line_2010}.
Starting from decoupled dot and leads we then time-evolve the system
for some time $\Gamma t$ until steady state has been reached. The
dot is initially empty. The coupling to the auxiliary leads described
in the previous chapter is $\eta=10^{-3}\Gamma$.

The effect of voltage on the spectral function is illustrated in Fig.~\ref{fig:voltage-effect}.
At zero voltage the Kondo peak can clearly be seen as it begins to
form (the temperature studied is at the upper edge of the Kondo regime).
With the application of a bias voltage, the peak lowers, widens and
eventually splits. While the magnitude of the Kondo effect decreases
when the system is driven away from equilibrium, the effect is obviously
not destroyed by the bias, and partial hybridization of the dot with
each lead occurs simultaneously. Except at frequencies much higher
than the bias, the spectral function is also significantly modified
by the nonequilibrium conditions, indicating that the equilibrium
spectral function is an inappropriate quantity for the description
of nonequilibrium physics. We note that the equilibrium aspects of
this problem may be addressed by numerical renormalization group\cite{costi_transport_1994},
which is expected to be more efficient at low temperatures. However,
outside equilibrium this has never been achieved, and it has been
suggested that this is due to a fundamental limitation of the Wilson
mapping\cite{rosch_wilson_2012}.

\begin{figure}
\includegraphics[width=8.6cm]{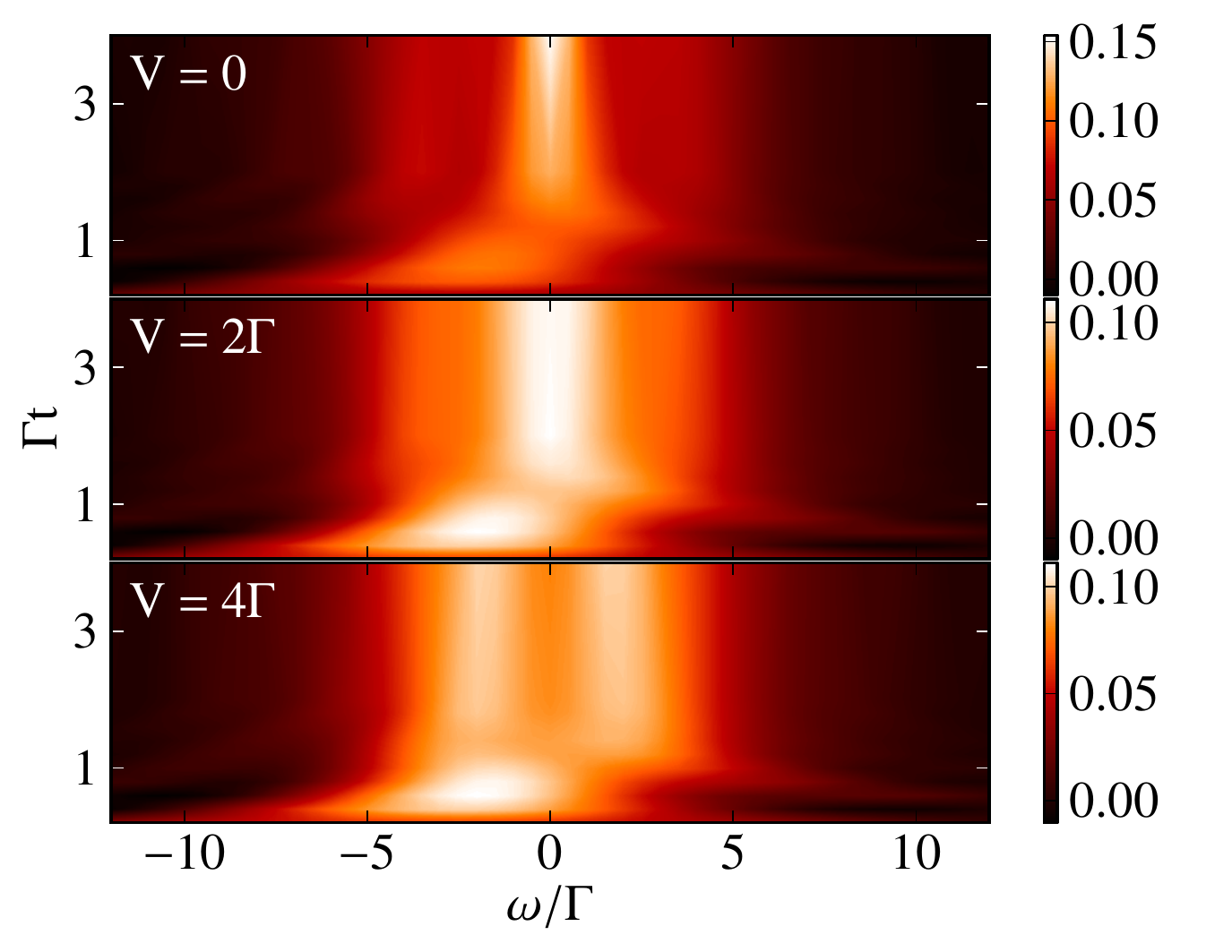}\caption{The time evolution of the spectral function $A_{\mathrm{aux}}\left(\omega\right)$
shown at several voltages, obtained from bold-CTQMC using the double-probe
auxiliary lead formalism, obtained at $\eta=10^{-3}\Gamma$.\label{fig:voltage-time-dependence}}
\end{figure}

The time dependence of $A_{\mathrm{aux}}$ is illustrated in Fig.~\ref{fig:voltage-time-dependence},
which shows what the results of a time-dependent measurement of $A_{\mathrm{aux}}$
would look like if the dot begins devoid of electrons and decoupled
from all leads. At short times, near the bottom edge of each of the
three panels, a peak forms (before disappearing) near the single-particle
resonance energy $\varepsilon_{\sigma}=-3\Gamma$. This corresponds
to the availability of electronic levels and is quickly followed by
the formation of corresponding hole levels at positive frequencies,
though experimentally we would not be able to distinguish electrons
from holes. At longer times, as the dot begins to fill, one can observe
near the top edges of the panels the formation of the steady-state
spectral properties including the equilibrium Kondo resonance (top
panel) and its voltage widened (middle panel) and split (bottom panel)
variations. The calculated quantity (and the hypothetical experiment)
therefore provides direct access not only about steady state spectral
properties, but also to the dynamical evolution of the system's total
density of single particle excitations. Combined with knowledge about
the initial conditions, this provides information about the population
dynamics.

\begin{figure}
\includegraphics[width=8.6cm]{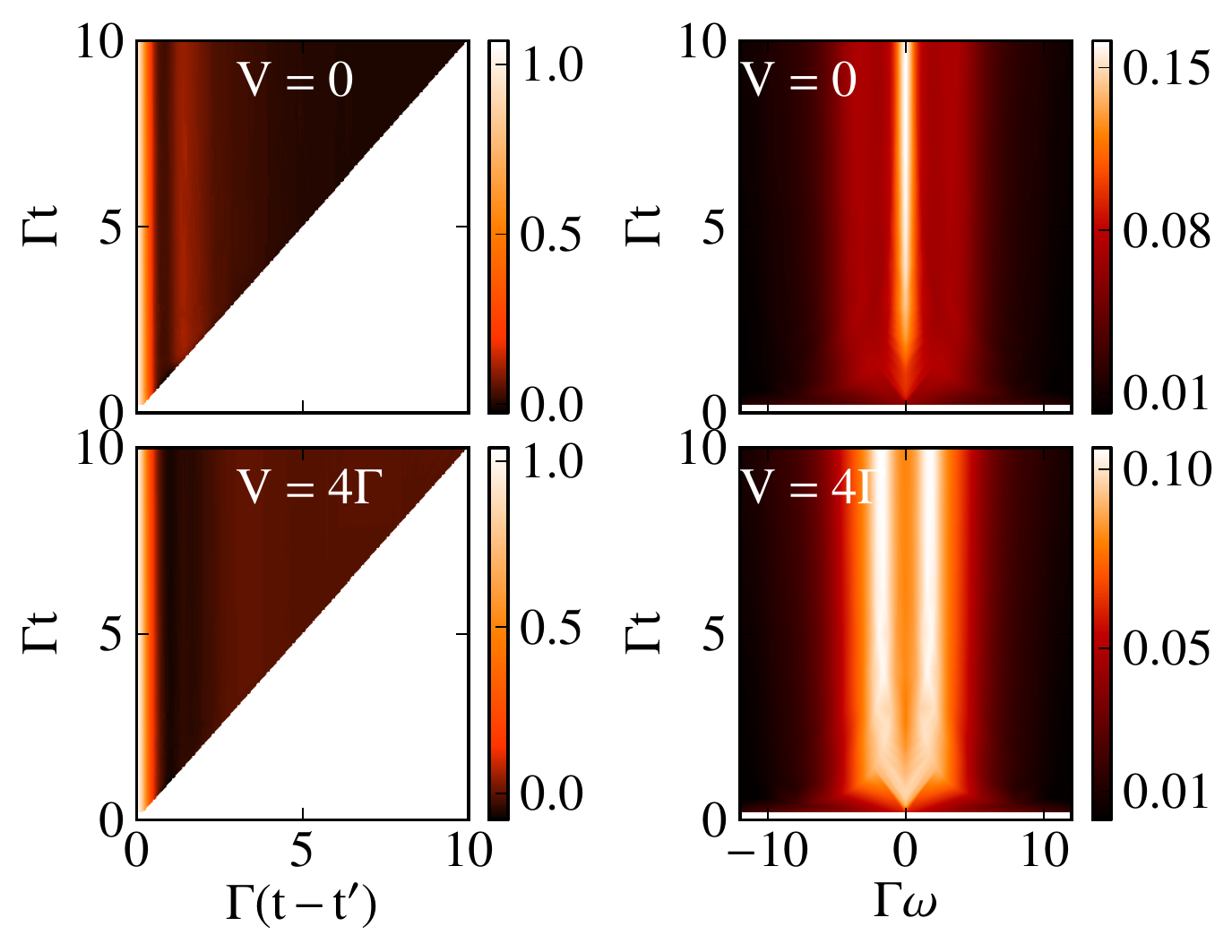}\caption{The time evolution of the real part of the retarded Green's function
$\Re\left\{ G^{r}\left(t,t-t^{\prime}\right)\right\} $ (left panels)
and the spectral function $A\left(\omega\right)$ (right panels) at
voltages indicated, as calculated from two time correlation functions
within bold-CTQMC.\label{fig:two_time_correlations}}
\end{figure}

A more conventional view of the dynamics is provided in Fig.~\ref{fig:two_time_correlations}.
Here we display the standard spectral function $A\left(\omega\right)$
at $V=0$ and $V=4\Gamma$ as a two-time correlation function (left
panels) and as a function of frequency and time (right panels). Notably,
these correlation functions are exactly the objects used in time-dependent
DMFT \cite{eckstein_thermalization_2009,eckstein_nonequilibrium_2010}.
The frequency-space property is obtained from the discrete Fourier
transform of the two-time property. The two-time correlations exhibit
little structure in the cases shown, due to the lack of explicit time
dependence in the Hamiltonian, but interestingly the nonequilibrium
nature of the dynamical evolution actually adds some noticeable correlations
at long times. At finite times frequency resolution is limited to
$\Delta\omega=\frac{\pi}{t}$, where $t$ is the propagation time.
In the finite voltage case, it actually seems as if a central peak
forms before the peak splitting occurs, which differs from what is
observed when $A_{\mathrm{aux}}$ is examined. Also, unlike $A_{\mathrm{aux}}$,
$A$ obeys particle--hole symmetry at all times.

\begin{figure}
\includegraphics[clip,width=8.6cm]{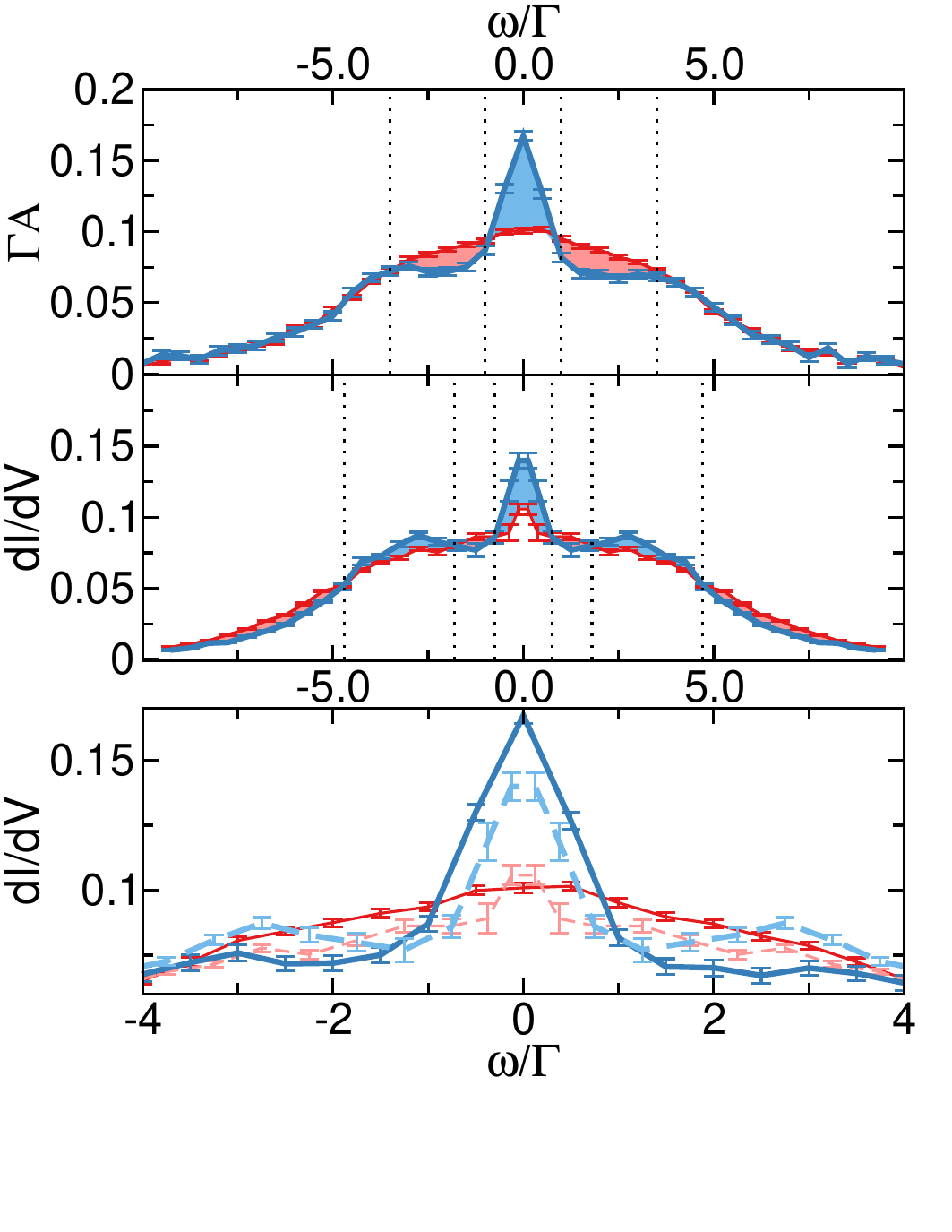}\caption{The steady state spectral function $A\left(\omega\right)$ (upper
panel) compared to the steady state differential conductance (middle
panel) as a function of half the voltage at two different inverse
temperatures $\beta$. Both observables are obtained from bold-CTQMC
at $\Gamma t=10$. The area between the curves are shaded according
to the maximal value and vertical dotted lines mark crossing points.
The bottom panel displays all results: the solid lines are the spectral
function and the dashed curves are the differential conductance.\label{fig:spectral_function_vs_current}}
\end{figure}

Within linear deviations from equilibrium in the voltage, the normal
differential conductance through the device---that is, the voltage
derivative of the current $I$ through the strongly coupled left or
right terminal---can be interpreted as an estimator for the equilibrium
spectral function. However, as Fig.~\ref{fig:voltage-effect} clearly
shows, the application of voltage beyond the linear response regime
significantly modifies the spectral density. It is therefore of some
interest to see how the use of normal current in the nonequilibrium
case fares in practice as a measure of equilibrium properties within
a numerically exact framework. The top panel of Fig.~\ref{fig:spectral_function_vs_current}
shows the equilibrium spectral function at two different inverse temperatures.
Below this, the lower panel of Fig.~\ref{fig:spectral_function_vs_current}
displays the steady state differential conductance for the same parameter
sets.

The two heavy blue $\beta\Gamma=3$ curves or the two lighter red
$\beta\Gamma=1$ curves appear superficially similar at first glance.
The clearest differences are a slight lowering and narrowing of the
Kondo peak, a slight accentuation of the Hubbard peaks, and a small
Kondo-like peak which appears in the differential conductance at a
temperature where it does not yet exist in the spectral function.
Comparing the two sets of curves side by side, however, brings the
inaccuracies of the differential conductance as an estimator for the
spectral function into sharp contrast: the differential conductance
exhibits a high-frequency temperature dependence completely absent
from the spectral function, with a temperature dependent trend at
intermediate frequencies that is actually reversed. A certain degree
of caution is therefore appropriate when applying the linear-response
interpretation to current measurements.

To summarize, we have implemented the computation of Green's functions
within real time bold-QMC in nonequilibrium using both correlation
functions and a double-probe auxiliary current formalism. We obtained
the spectral function of the nonequilibrium Anderson model and demonstrated
the voltage splitting of the Kondo peak within a general, numerically
exact framework. Through our formalism the dynamics of the excitation
density of states starting with a coupling or gate quench and up to
the formation of a Kondo peak was studied, with and without a bias
voltage (the formalism is also applicable to other quench types, such
as voltage, temperature or interaction quenches). We have shown that
the auxiliary lead interpretation and the associated experimental
setup provides access not only to steady state spectral properties,
but also to information about the excitation and (indirectly) population
dynamics of the system. Finally, we have discussed the use of current
measurements in the more common two lead setup to access the equilibrium
spectral properties of the Anderson model, demonstrating that while
the differential conductance provides a good qualitative estimator
for spectral functions, it also fails in reproducing temperature trends
at lower frequencies while introducing spurious trends at high ones.

Looking forward, the tools presented here not only provide new insight
into transport in quantum impurity models, but also provide the functionality
required by an impurity solver within nonequilibrium DMFT: bold-CTQMC
can provide spectral data which can be incorporated into DMFT calculations
incorporating multiple leads at different thermodynamic parameters,
as well as two time correlation functions for time-dependent DMFT.
It is practically useful up to times and interaction strengths substantially
greater than those of previous Monte Carlo methods, while maintaining
Monte Carlo's critical advantage in resolution over other methods.
The bold-CTQMC method is therefore expected to have important consequences
in the study of strongly correlated systems out of equilibrium. 
\begin{acknowledgments}
The authors would like to thank Rainer Härtle and Eran Rabani for
their many insightful comments. GC is grateful to the Yad Hanadiv--Rothschild
Foundation for the award of a Rothschild Postdoctoral Fellowship and
NSF DMR 1006282. GC and EG acknowledge TG-DMR120085 and TG-DMR130036
for computer time. DRR acknowledges NSF CHE-1213247. AJM acknowledges
NSF DMR 1006282. EG acknowledges DOE ER 46932. Our implementations
were based on the ALPS \cite{bauer_alps_2011} libraries. 
\end{acknowledgments}
\bibliographystyle{apsrev4-1}
\bibliography{Library}

\end{document}